\documentclass[%
 aip,
 amsmath,amssymb,
 reprint,%
]{revtex4-1}

\usepackage{graphicx}
\usepackage{dcolumn}
\usepackage{booktabs} 
\usepackage{bm}

\usepackage[utf8]{inputenc}
\usepackage[T1]{fontenc}
\usepackage{mathptmx}
\usepackage{etoolbox}

\makeatletter
\def\@email#1#2{%
 \endgroup
 \patchcmd{\titleblock@produce}
  {\frontmatter@RRAPformat}
  {\frontmatter@RRAPformat{\produce@RRAP{*#1\href{mailto:#2}{#2}}}\frontmatter@RRAPformat}
  {}{}
}%
\makeatother
\begin{document}

\preprint{AIP/123-QED}

\title{Realization of two-qubit gates and multi-body entanglement states in an asymmetric superconducting circuits}
\author{Tao Zhang}
 \affiliation{College of Science, Hangzhou Dianzi University.}
\author{Chaoying Zhao}
\email{zchy49@163.com}
\affiliation{College of Science, Hangzhou Dianzi University, Hangzhou 310018, China}
\affiliation{State Key Laboratory of Quantum Optics and Quantum Optics Devices, Institute of Opto-Electronics, Shanxi University, Taiyuan 030006, China}

\date{\today}

\begin{abstract}
In recent years, the tunable coupling scheme has become the mainstream scheme for designing superconducting quantum circuits. By working in the dispersive regime, the ZZ coupling and high-energy level leakage can be effectively suppressed and realize a high fidelity quantum gate. We propose a tunable fluxonium-transmon-transmon (FTT) coupling scheme. In our system, the coupler is a frequency tunable transmon qubit. Both qubits and coupler are capacitively coupled. The asymmetric structure composed of fluxonium and transmon will optimize the frequency space and form a high fidelity two-qubit quantum gate. By decoupling, the effective coupling strength can be easily adjusted to close to the net coupling between qubits. We numerical simulation the master equation to reduce the quantum noise to zero. We study the performance of this scheme by simulating the general single-qubit $\left(X_{\pi/2}\right)$ gate and two-qubit $\left(iSWAP\right)$ gate. In the bias point of the qubits, we achieve a single qubit gate with 99.99$\%$ fidelity and a two-qubit gate with 99.95$\%$ fidelity. By adjusting the nonlinear Kerr coefficient of fluxonium to an appropriate value, we can achieve a multi-body entanglement state. We consider the correlation between the two qubits and the coupler, and the magnetic flux passing through one qubit has an effect on the other qubit and the coupler. Finally, we analyze the quantum correlation of the two-body entanglement state. 
\end{abstract}

\maketitle



\section{\label{sec:level1}INTRODUCTION}
Quantum superconducting circuits based on transmon have become a mainstream trend over the years \cite{Majer2007,dicarlo2009,zeytinouglu2015,egger2019,krinner2020}, because of transmon's unique circuit structure \cite{koch2007}, the charge dispersion is small and insensitive to the charge noise. At the same time, transmon can maintains a good non-harmonic so that the operation can be accurately addressed. However, a high fidelity quantum gates require coherent qubits and suitable qubit-qubit coupling \cite{preskill2018}. As the circuit structure becomes more and more complex, unnecessary qubit interactions will degrade gate performance. A coupling transmon can deal with unnecessary coupling and frequency congestion \cite{mckay2016,mundada2019,li2020,han2020,xu2020}.The effective coupling between two qubits can be controlled by adjusting the intermediate-mediated coupler frequency. The coupling strength is set to zero when we running a single qubit gate, and the coupling between qubits can be turned on when we running a double qubit gate.

The selection of fluxonium as a qubit for superconducting quantum circuits is a promising improvement \cite{xiang2013,arute2019,place2021,jurcevic2021}. Fluxonium can prevents charge and flux-induced dephase by shunting the Josephson junction and a large linear inductor \cite{bao2022}, while retaining a large an-harmonies. Furthermore, in order to achieve a considerable coherence time \cite{nguyen2019}, the fluxonium needs to work at a much lower frequency than that of the transmon. The energy relaxation can be suppressed at a low qubit frequency, which means a reduction in the two-qubit gate velocity limited by the directly qubit-qubit coupling strength. Fluxonium has been used in tunable coupling systems \cite{moskalenko2021}, enabling flux quantum circuits with additional modes via capacitive coupling of quantum bits and couplers.

Imperfect two-qubit gate operation is an urgent problem in superconducting quantum circuits. In order to improve the fidelity of quantum gates, different experiments research groups choose different types of qubits or coupler. Neill et al. achieved a high coherence and fast tunable coupling by using Xmon qubits \cite{chen2014}, and a two-qubit idle gate with 99.56$\%$ fidelity. Hayato et al. implemented a high-performance two-qubit parametric gate based on double-transmon couplers \cite{kubo2023}. Although the fidelity is high enough, the detuning between qubits is not large enough and the qubit frequency is high. Therefore, in order to improve the detuning between qubits and make the system work in dispersive regime, we choose fluxonium as a qubit to couple with transmon, which can alleviate the problem of frequency-crowding \cite{Hertzberg2020}.

In this work, in order to solve the problem of frequency-crowding and form a stable and reliable quantum gate, we propose a fluxonium and transmon systems with tunable coupler capacitance coupling(FTT), which can achieve a high fidelity two-qubit gates. By adjusting the frequency of the coupler and changing the effective coupling strength between the two qubits, the net coupling can be effectively closed. At this time, we can simulate a single-qubit gate in our system without interference from other qubits. Two-qubit gates  operate in the dispersive regime can suppress leakage of higher excited states. In this way, the coherence time of two qubits and the fidelity of the system can be improved. We can simulate their relaxation time and dephasing time in scqubits, and compare the relationship between the coherence time and the gate time. In addition, we use an asymmetric structure, and realize the preparation of entanglement states and analyze the quantum correlation of the two-body entanglement states based on the nonlinear Kerr coefficient of fluxonium.

\section{The Theoretical Model}
\begin{figure}
\includegraphics[width=0.49\textwidth]{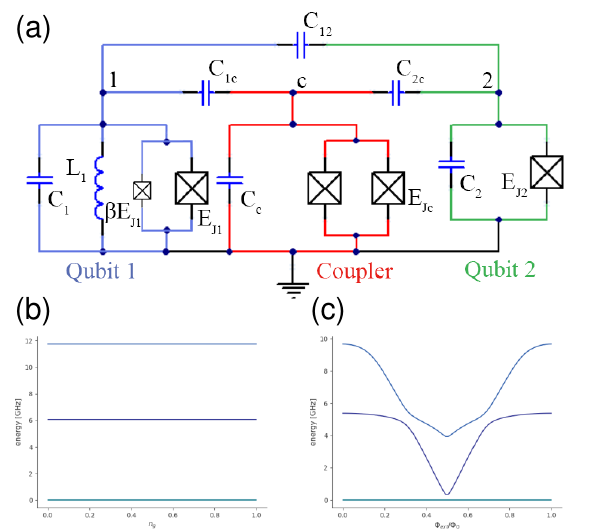}
\caption{\label{fig:1} (a)The equivalent circuit diagram. Each circuit has different colors: fluxonium(blue), transmon(green), and coupler(red). (b)The three levels of transmon qubit as a function of the effective offset charge. (c)The three levels of fluxonium qubit as a function of the external magnetic flux.}
\end{figure}

We use a nonlinear fluxonium qubit and a transmon qubit, which are capacitively coupled to a grounded transmon coupler, as shown in Fig.~\ref{fig:1}(a). The tunable coupler consists of a transmon with an tunable frequency, and there is a direct capacitive coupling between the fluxonium and transmon qubits, the direct coupling between two qubits is much weaker. We use a Josephson loop (superconducting quantum interference device) instead of the Josephson junction in fluxonium, which can produce nonlinear effect and the frequency of transmon is fixed. When we only consider the qubits model and the capacitance coupling between them, the Hamiltonian of our system is (see Appendix A)
\begin{equation}
\begin{split}
\hat{H}=&4E_{c_1}\hat{n}_1^2-\beta E_{J_1}\cos\varphi_1-E_{J_1}cos\left(\varphi_{ext}-\varphi_1\right)\\&+\frac{1}{2}E_{L_1}\hat{\varphi}_1^2+4E_{c_2}\hat{n}_2^2-E_{J_2} cos{\hat{\varphi}_2}+4E_{cc}\hat{n}_c^2\\&-E_{J_c}\cos{\hat{\varphi}_c}+g_{1c}\hat{n}_1\hat{n}_c+g_{2c}\hat{n}_2\hat{n}_c+g_{12}\hat{n}_1\hat{n}_2,
\end{split}
\end{equation}

where $E_C$, $E_J$, and $E_L$ represent the charging, Josephson, and inductive energies, respectively. The ratio of Josephson energy of the small and big junction is $\beta$=0.1. Subscripts $i=1,2$ index the fluxonium nodes and the transmon nodes, respectively, and subscript $c$ labels the coupler node. The coupling strength $g_{ij}$ (i,j$\in$ 1,2,c) represents the coupling strength between different modes, and the direct coupling strength $g_{12}$ is far less than the indirect coupling strength $g_{1c}$ and $g_{2c}$. 

In our numerical simulation, the Fluxonium and transmon are used at the zero flux bias point ($\Phi_{ext}$/$\Phi_0=0$) and the charge bias point ($n_g=0.5$), respectively, as shown in Fig.~\ref{fig:1}(b) and Fig.~\ref{fig:1}(c). The external magnetic flux can be applied to the Josephson junction loop in the coupler through the magnetic flux-bias line, thus wen can change the frequency of the tunable coupler. 
The transmon with charging energy $E_C=0.32GHz$ and inductance energy $E_J=16.0GHz$, the transmon needs to work under the limit of $E_C<E_J$. The fluxonium is very sensitive to the external magnetic flux, when we simulate a two-qubit gate, we set the external flux to zero. When we adjust the nonlinear Kerr coefficient, we need to turn on it. 

\begin{figure}
\includegraphics[width=0.5\textwidth]{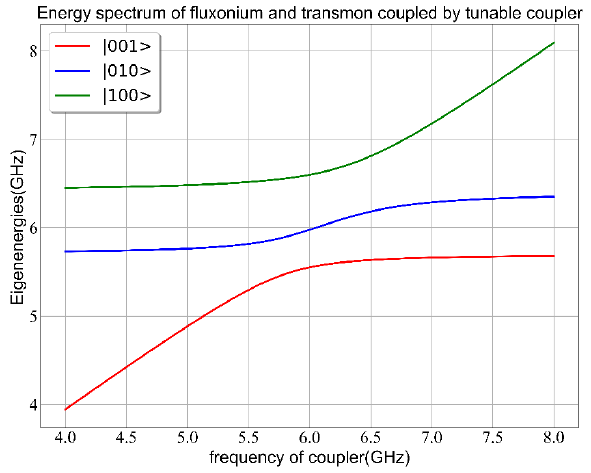}
\caption{\label{fig:2} Spectrum of fluxonium and transmon coupled to a tunable coupler with different local magnetic flux for the coupler. There are two avoided crossing exists: $|100\rangle\leftrightarrow|010\rangle$ and $|001\rangle\leftrightarrow|010\rangle$. The computational state transitions formed by the effective Hamiltonian, the effective Hamiltonian can perform a iSWAP gate. }
\end{figure}

We describe the quantum state by using
the notation $|Q_1,C,Q_2\rangle$, where $Q_1$, $C$, and $Q_2$ denote the energy
eigen-states in the uncoupled basis of fluxonium, the
coupler, and transmon, respectively. We need to avoid crossing between computational state $|100\rangle$ and $|010\rangle$, we also need to avoid crossing between computational state$|001\rangle$ and $|010\rangle$. The size is twice the corresponding coupling strength as shown in Fig.~\ref{fig:2}.

The circuit parameters and the corresponding coupling strength are used in our numerical simulation as shown in Table ~\ref{tab:table1}. The qubit and coupler are detuned and the coupling is dispersive. The coupling of two qubits has two channels, in which direct coupling through capacitance and indirect coupling through coupler. The coupler decouple by the Schrieffer-Wolff transformation(SWT)\cite{yan2018}. Finally, the effective Hamiltonian of two qubits can be obtained (see Appendix B):

\begin{equation}
\begin{split}
\hat{\tilde{H}}=&\tilde{\omega}_1\hat{a}_1^\dagger\hat{a}_1+K\hat{a}_1^\dagger\hat{a}_1^\dagger\hat{a}_1\hat{a}_1+\tilde{\omega}_2\hat{a}_2^\dagger\hat{a}_2\\&+\frac{\tilde{\alpha_2}}{2}\hat{a}_2^\dagger\hat{a}_2^\dagger\hat{a}_2\hat{a}_2+\tilde{g}(\hat{a}_1^\dagger\hat{a}_2+\hat{a}_1\hat{a}_2^\dagger),
\end{split}
\end{equation}

where $\tilde{\omega_1}$ and $\tilde{\omega_2}$ are the frequencies of fluxonium and transmon, respectively. $K$ is the nonlinear Kerr coefficient of fluxonium. $\tilde{\alpha_2}$ is the anharmonicity of transmon, and $\tilde{g}$ is the effective coupling strength between fluxonium and transmon. $\hat{a}_m(\hat{a}_m^\dagger)$ corresponds to the generation (annihilation) operator of the corresponding mode.

\begin{figure}
\includegraphics[width=0.5\textwidth]{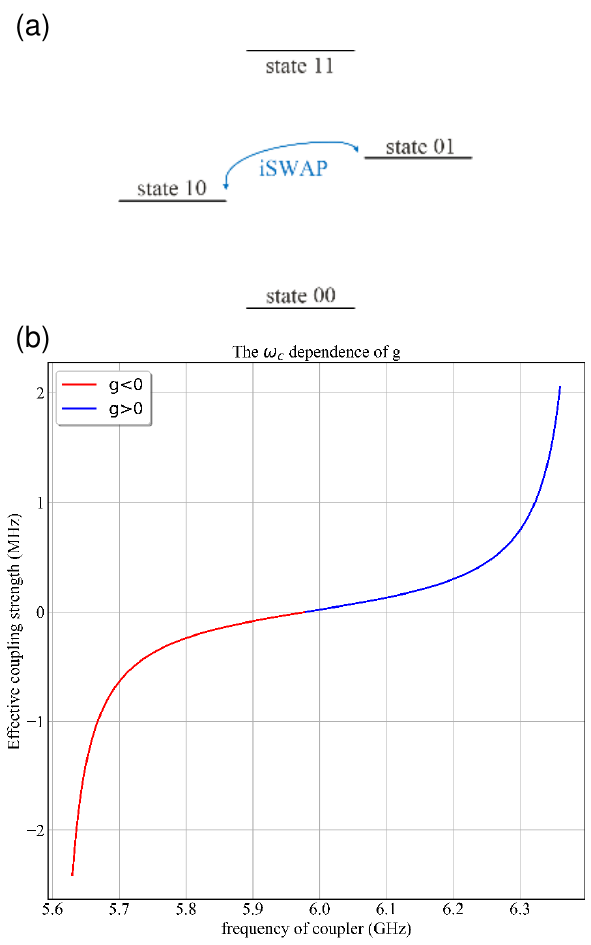}
\caption{\label{fig:3} (a)Schematic of the level structure of FTT system with indication of the transitions utilized in the $iSWAP$ gate. The $iSWAP$ gate connects $|10\rangle$ and $|01\rangle$. By setting the gate time, the states of qubits exchange and attaches with a global phase. As shown in the figure are computational states, while the  higher energy levels without mark in the figure are non-computational states. (b)The 
frequency $\omega_c$ of the coupler depends on the effective coupling strength $2\tilde{g}$. The coupler frequency is 5.97$GHz$, which corresponds to the coupling zero point. The coupling strength of the red part is less than zero, and the coupling strength of the blue part is greater than zero, both of which can be adjusted through the coupler frequency.}
\end{figure}

In Fig.~\ref{fig:3}(a), we preset the energy level diagram of the tunable coupling system and the range of transitions. Through the $iSWAP$ gate, we can realize the transition of $|01\rangle\leftrightarrow|10\rangle$. By applying microwave pulses to different qubits, we can realize the transition from state $|00\rangle$ to state $|01\rangle$ or 
state $|10\rangle$.

\begin{table}
\caption{\label{tab:table1}Circuit system parameters and the corresponding coupling strength are used in numerical simulation.}
\begin{tabular}{|c|c|c|c|c|}
\toprule [1pt]
&$E_C(GHz)$&$E_L(GHz)$&$E_J(GHz)$&$\omega/2\pi(GHz)$\\
\midrule [1pt]
Fluxonium & 0.9 & 1.0 & 4.5 & 5.7\\ \hline
transmon & 0.32 &  & 16.0 & 6.4\\ \hline
coupler & 0.32 &  & 12.8 & \\ \hline
\multicolumn{2}{|c|}{}&$g_{1c}(MHz)$&$g_{2c}(MHz)$&$g_{12}(MHz)$\\ \hline
\multicolumn{2}{|c|}{coupling strength} & 242.9 & 307.11 & 34.7\\
\bottomrule [1pt]
\end{tabular}
\end{table}

The effective coupling strength depends on the frequency of the coupler, as shown in Fig.~\ref{fig:3}(b). We choose the coupler frequency from 4.0$GHz$ to 8.0$GHz$ corresponds to the adjustment range of the tunable coupler. 

When the coupler frequency reaches to 5.97$GHz$, the net coupling between qubits need to completely turn off. When the whole system works at the coupling zero point, we apply microwave drive the state of a single qubit rotate accordingly without affect the state of another qubit. When we operate the two-qubit gates, we can open the interaction and realize the controllable gate operation. We can also change the frequency of the coupler by applying the magnetic flux pulse, so as to achieve the effect of changing the effective coupling strength.

The operation of quantum gates can produce leakage state\cite{chu2021}, the most important is the unnecessary excitation of the coupler, which makes the whole system transition to a non-computational state, and forming a gate error. We operate the coupler in the dispersion regime to perform the two-qubit gate operation, which strongly inhibits the leakage of the coupler to the excited state. In the dispersion limit, the frequency difference $\Delta_j=\omega_j-\omega_c$ between the coupler and the qubit is much larger than the coupling strength  $g_{jc}\ll|\Delta_j|(j=1,2)$.

\section{GATE PRINCIPLES}

\subsection{Single-qubit gates}

We simulate the quantum gates with QuTip in PYTHON\cite{johansson2012}. First, we set the effective coupling is zero and close the net coupling of two qubits. At this time, a flux pulse passing through the flux bias line. After turning off the coupling between two qubits, a microwave drive with the frequency of single qubit. The drive time of microwave determines the rotation angle of the quantum state on the Bloch sphere. Taking the rotating wave approximation, the driving Hamiltonian can be written as\cite{krantz2019}
\begin{equation}
\begin{split}
\tilde{H}_d=-\frac{\Omega}{2}V_0s(t)\left(I\sigma_x+Q\sigma_y\right),
\end{split}
\end{equation}

Where $\Omega$ is a dimensionless parameter determined by circuit parameters and $s(t)$ is a dimensionless envelope function. In addition, $V_0$ is the voltage amplitude, $\sigma_x$ and $\sigma_y$ are the Pauli matrices in the $x$ and $y$ directions, respectively. The in-phase component $I=\cos(\phi)$ and the out-of-phase component $Q=\sin(\phi)$ can control the axis of rotation. When $\phi=0$, the rotation revolves around the $x$-axis, and when $\phi=\frac{\pi}{2}$, the rotation revolves around the $y$-axis. We numerically simulate the $X_{\frac{\pi}{2}}$ gate, whose matrix expression is

\begin{equation}
\begin{split}
X_{\frac{\pi}{2}}=\frac{\sqrt{2}}{2}\left[\begin{array}{cc}
1 & -i\\
-i & 1\\
\end{array}\right]
\end{split}
\end{equation}

\begin{figure}
\includegraphics[width=0.5\textwidth]{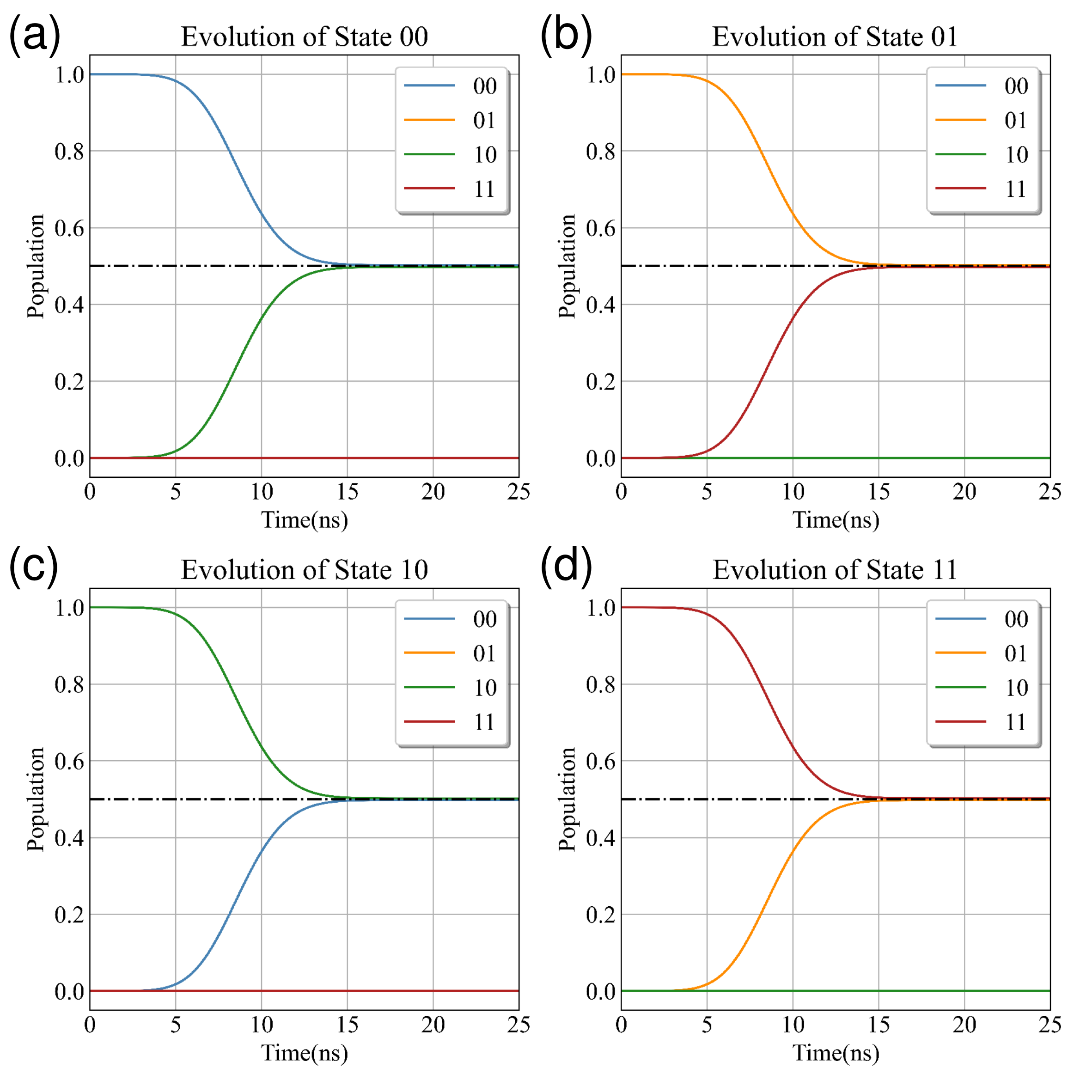}
\caption{\label{fig:4}Time evolution of populations for four initial computational states during the single-qubit gate: (a) The initial state is $|00\rangle$. (b) The initial state is $|01\rangle$. (c) The initial state is $|10\rangle$. (d) The initial state is $|11\rangle$. The horizontal black dotted line in the figure represents the population of $0.5$.}
\end{figure}

We simulate the $X_{\frac{\pi}{2}}$ gate on fluxonium by using modulated Gaussian pulses (with "mesolve" in QuTip). By applying a microwave pulse to realize a single-qubit gate, we simulate the evolution of the system state under four different initial states. The evolution of each state under different initial states are shown in Fig.~\ref{fig:4}.

We simulate the evolution of quantum states with an initial state of $|Q_1Q_2\rangle=|00\rangle,|01\rangle,|10\rangle$ and $|11\rangle$, respectively. We set the modulate pulse frequency $\omega_d$ as the frequency of fluxonium $\omega_1=5.7$GHz and the gate time as $15ns$. From the figure, we can see that $|0\rangle\xrightarrow[]{X_{\pi/2}}\frac{\sqrt{2}}{2}(|0\rangle-i|1\rangle)$ and $|1\rangle\xrightarrow[]{X_{\pi/2}}\frac{\sqrt{2}}{2}(-i|0\rangle+|1\rangle)$, which is consistent with the fidelity theoretical result. Due to the pulse acting on fluxonium, the quantum state of transmon remains unchanged. 

\begin{equation}
\begin{split}
F(\rho,\sigma)=tr\sqrt{\rho^{1/2}\sigma\rho^{1/2}}
\end{split}
\end{equation}

Finally, we use fidelity to calculate the proximity between actual quantum state $\rho$ and ideal quantum state $\sigma$. The fidelity of the single-qubit gate is 99.99$\%$, and the error occurs after six decimal places. 

Thus, we obtain a single-qubit gate with high fidelity by modulating the pulse waveform. If it is necessary to obtain a state with a better fidelity, which may be necessary to optimize the pulse waveform to further suppress the leakage of excited states.

\subsection{Two-qubit gates}

\begin{figure}
\includegraphics[width=0.5\textwidth]{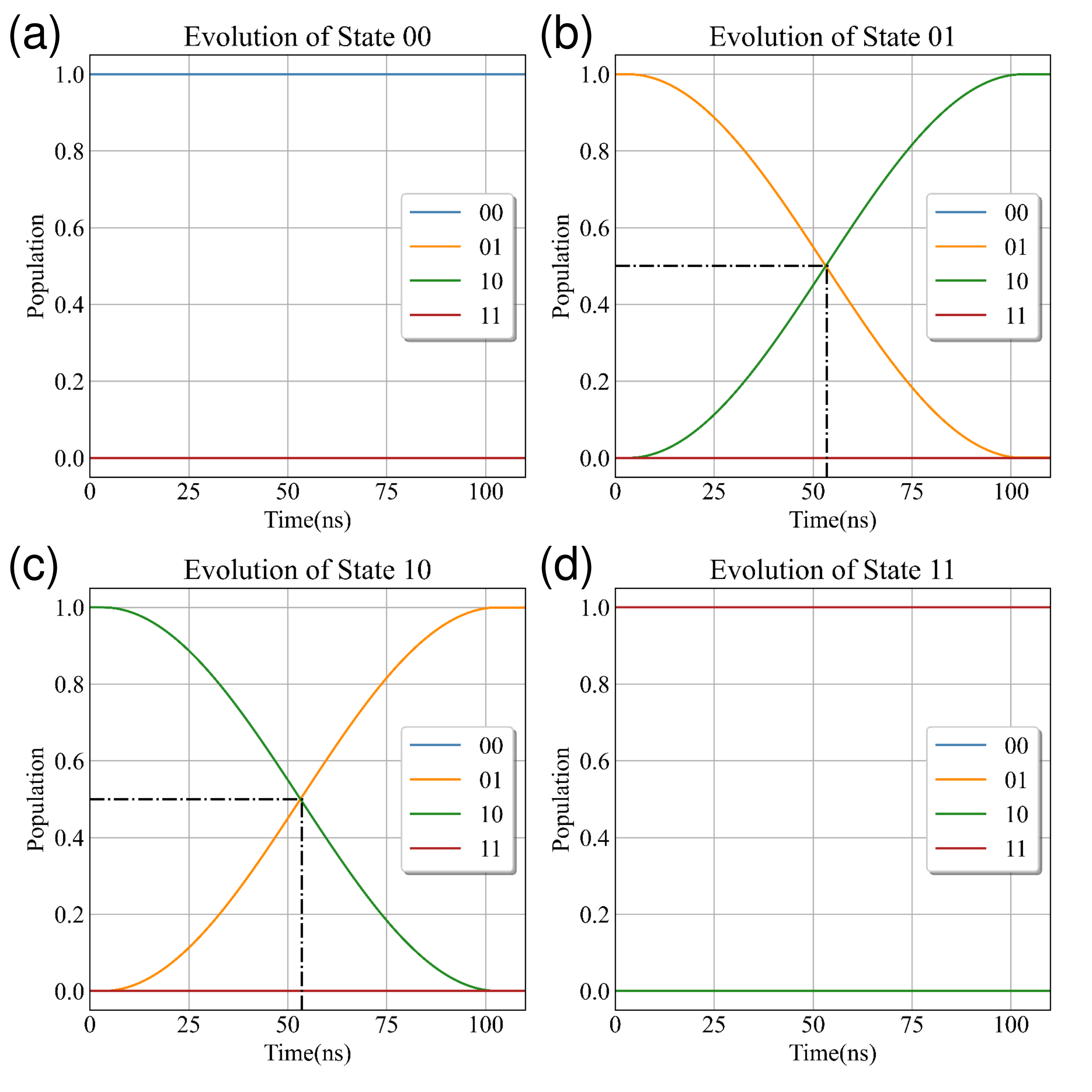}
\caption{\label{fig:5}Time evolution of populations for four initial computational states during the two-qubit gate: (a) The initial state is $|00\rangle$. (b) The initial state is $|01\rangle$. (c) The initial state is $|10\rangle$. (d) The initial state is $|11\rangle$. The horizontal black dotted line represents the population of $0.5$, and the vertical dotted line represents the time.}
\end{figure}

Next, we implement a universal two-qubit gate via an effective Hamiltonian, and only consider the first two energy levels of the qubits. When fluxonium and transmon are in resonance and the two qubits are truncated to two enegry levels, by the rotating wave approximation, the Hamiltonian between qubits approximates to

\begin{equation}
\begin{split}
\hat{H}_{qq}=\tilde{g}(\sigma_1^+\sigma_2^-+\sigma_1^-\sigma_2^+)
\end{split}
\end{equation}

The unitary corresponding to the qubit-qubit interaction is

\begin{equation}
\begin{split}
U_{qq}(t)=e^{-i\hat{H}_{qq}t}=\left[\begin{array}{cccc}
1 & 0 & 0 & 0 \\
0 & \cos(gt) & -i\sin(gt) & 0\\
0 & -i\sin(gt) & \cos(gt) & 0\\
0 & 0 & 0 & 1 
\end{array}\right]
\end{split}
\end{equation}

When two qubits are in resonance and the gate time $t=\frac{\pi}{2g}$, we can get the matrix expression of $iSWAP$ gate:

\begin{equation}
\begin{split}
iSWAP=U_{qq}(\frac{\pi}{2g})=\left[\begin{array}{cccc}
1 & 0 & 0 & 0 \\
0 & 0 & -i & 0\\
0 & -i & 0 & 0\\
0 & 0 & 0 & 1 
\end{array}\right]
\end{split}
\end{equation}

Therefore, we can use the pulse sequence to realize the $iSWAP$ gate: (1)an error type square wave acting on fluxonium or transmon to make up the frequency difference between two qubits, in order to reaches to the resonance state. (2)A rectangular magnetic flux pulse acting on the coupler will changes the frequency of the coupler so that the effective coupling strength reaches to the $iSWAP$ gate operating point. The time length of the two pulses is $t=\frac{\pi}{2g}$, which can make the $iSWAP$ gate operate completely and prevent the leakage of other quantum states.

\begin{figure}
\includegraphics[width=0.5\textwidth]{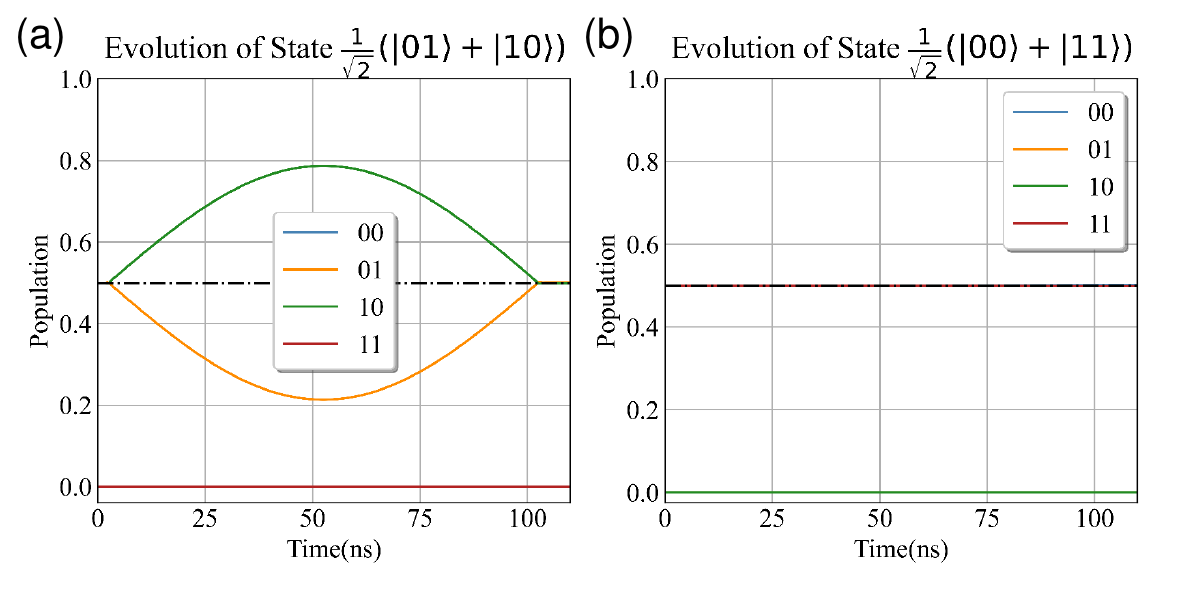}
\caption{\label{fig:6}Time evolution of populations for two special initial computational states in the during of the two-qubit gate: (a) The initial state is $\frac{1}{2}\left(|01\rangle+|10\rangle\right)$. (b) The initial state is $\frac{1}{2}\left(|00\rangle+|11\rangle\right)$. }
\end{figure}

By applying a magnetic flux pulse and a rectangular pulse to realize the two-qubit gate, we simulate the evolution of system states under six different initial states. We simulate the $iSWAP$ gates with QuTip in PYTHON. The evolution of each state under different initial states as shown in Fig.~\ref{fig:5}. We simulate the evolution of quantum states with an initial state of $|Q_1Q_2\rangle=|00\rangle,|01\rangle,|10\rangle$ and $|11\rangle$, respectively. In this case, the total qubit-qubit coupling is 250$MHz$ and the coupler frequency is tuned to 4.27$GHz$. We can see that $|01\rangle\xrightarrow[]{iSWAP}-i|10\rangle$ and $|10\rangle\xrightarrow[]{iSWAP}-i|01\rangle$, which is consistent with the fidelity theoretical results. After exchanging the states of two qubits will result in a global phase of $\frac{\pi}{2}$. When the initial state is at $|00\rangle$ or $|11\rangle$, it is a horizontal straight line. We get the fidelity of the two quantum states is 99.97$\%$.

In order to show the quantum evolution of some special initial states, we select the evolution of two Bell States under the $iSWAP$ gate, as shown in Fig.~\ref{fig:6}. The quantum state with an initial state of $\frac{1}{\sqrt{2}}(|01\rangle+|10\rangle)$ exchanges internally under the action of the $iSWAP$ gate and finally returns to the initial state, there is an additional global phase. However, the quantum state with initial state $\frac{1}{\sqrt{2}}(|00\rangle+|11\rangle)$ does not change ecologically through the action of $iSWAP$ gate, there are two straight lines. We finally get the result of 97.63$\%$. A gate operation can be completed within the coherence time of two qubits.



\begin{figure}
\includegraphics[width=0.48\textwidth]{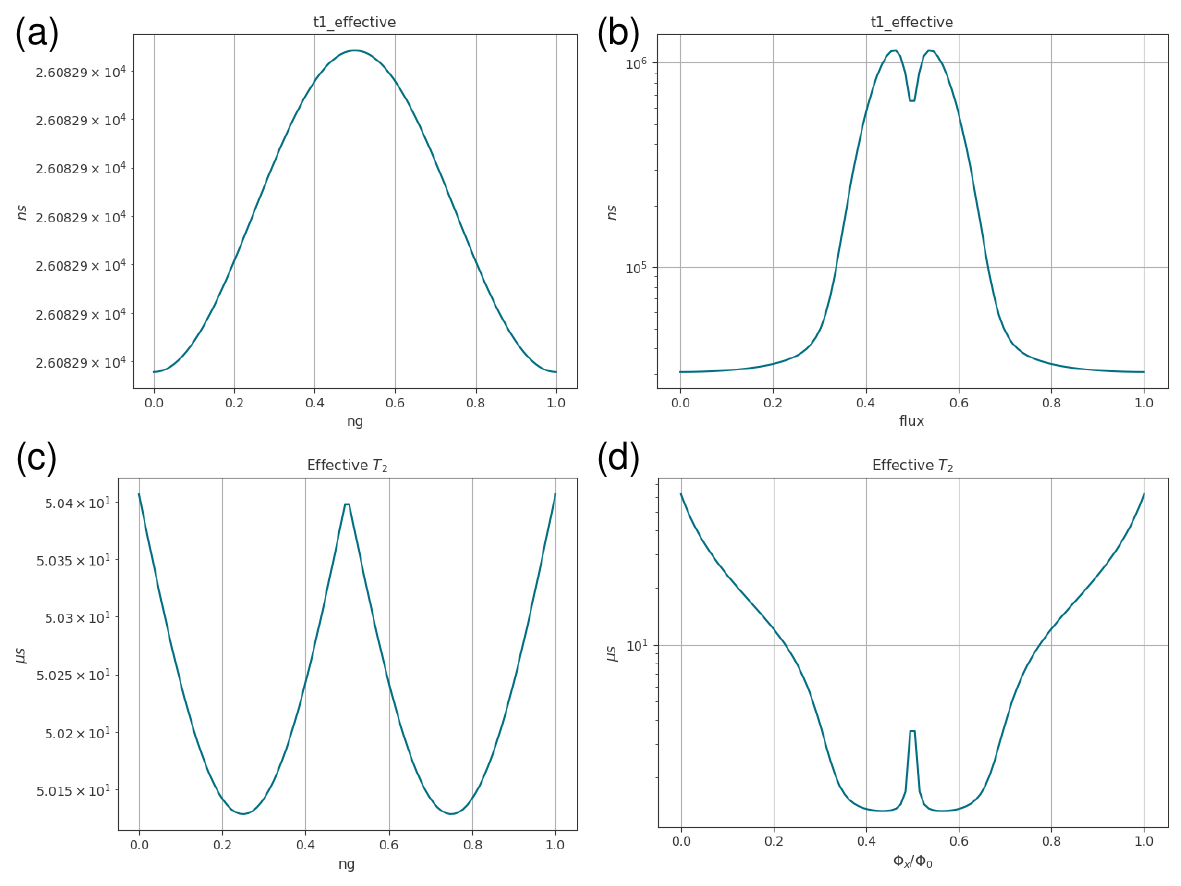}
\caption{\label{fig:7} (a) The effective coherence time of transmon. (b) The effective coherence time of fluxonium. (c) The effective dephase time of transmon. (d) The effective dephase time of fluxonium.}
\end{figure}

\begin{table}
\caption{\label{tab:table2}Effective coherence time and dephase time of fluxonium and transmon, and single-qubit gate and two-qubit gate time.}
\begin{tabular}{|c|c|c|}
\toprule [1pt]
& coherence time$T_1(ms)$ & dephase time$T_2(ms)$\\
\midrule [1pt]
Fluxonium & 309.42 & 56.7 \\ \hline
transmon & 260.82 & 506.99 \\ \hline
& single-qubit gate$(ns)$ & two-qubit gate$(ns)$ \\ \hline
{gate time} & 15 & 103 \\
\bottomrule [1pt]
\end{tabular}
\end{table}

Depending on the physical structure of fluxonium, capacitance and inductance are introduced into charge and inductor noise channels, respectively. Because of the SQUID in fluxonium, we also take into account the flux bias channel connected to it, as well as the quasi-particle noise introduced by the Josephson junction. Whereas, for a fixed-frequency transmon, we only consider the charge noise caused by its capacitance.
 
Here we use Python's scqubits\cite{groszkowski2021} package to simulate the three energy levels and coherence time of two qubits. The qubit parameters used in our simulation are shown in Table ~\ref{tab:table1}. For transmon, the offset charge we selected from 0 to 1; For fluxonium, we choose the external magnetic flux from 0 to $\Phi_0$. When two qubits work at the bias point, the frequency of fluxonium is $5.7GHz$, and the frequency of transmon is $6.4GHz$. From Fig.~\ref{fig:7}, we can get the effective coherence time $T_1$ of fluxonium is 309.42 $ms$, and the effective coherence time $T_1$ of the transmon is 260.82 $ms$. In addition, the effective dephase time $T_2$ of fluxonium is 56.7 $ms$, and the effective dephase time $T_2$ of transmon is 506.99 $ms$. The operation time of single-qubit gate and two-qubit gate is 15 $ns$ and 103 $ns$, respectively. Therefore, from the perspective of theoretical calculation, we can complete the whole gate operation in the finite qubit coherence time.

\section{Preparation of Schrödinger Cat states}

We can prepare the non-classical quantum states by using Kerr non-linearity effect, we need to take into account the nonlinear effects of the fluxonium. First, fluxonium can prepare the coherent state $|\pm\alpha\rangle$ by using the coherent pulses. The coherent state will evolve with the Kerr non-linearity effects. Then, after the displacement pulse $D(\alpha)$, the Kerr non-linearity can be adjusted by using the fluxonium SQUID's magnetic flux bias pulse. Under the influence of the nonlinear Hamiltonian, the rotating speed of the coherent state in the phase space is not uniform, so the kerr Cat state can be produced at the right time. The evolution of a qubit with the initial state $|\alpha\rangle$ under Kerr non-linearity effect can be written as \cite{he2023fast}

\begin{equation}
\begin{split}
|\psi(t)\rangle&=e^{\frac{i}{\hbar}Ht}|\alpha\rangle=e^{\frac{i}{2}K(\hat{a}^\dagger\hat{a})^2t}|\alpha\rangle
\\&=e^{-\lvert\alpha\rvert^2/2}\sum\limits_n\frac{\alpha^n}{\sqrt{n!}}e^{i\frac{K}{2}n^2t}|n\rangle,
\end{split}
\end{equation}

The coherent state is expanded into a Fock state for reforming calculation, and when $Kt$ is at a certain value, the final state will evolve into a multi-body Cat state \cite{Tan2023,Zhao2019,Zhao2020}. In the calculation, the relationship between Kerr non-linearity and time that separates the real part from the imaginary part is $t=\tau_0/m$ and $\tau_0=2\pi/K$. In particular, when $t=\pi/K$, the final state is

\begin{equation}
\begin{split}
|\psi(t)\rangle=|\alpha\rangle\pm i|-\alpha\rangle,
\end{split}
\end{equation}

\begin{figure}
\includegraphics[width=0.48\textwidth]{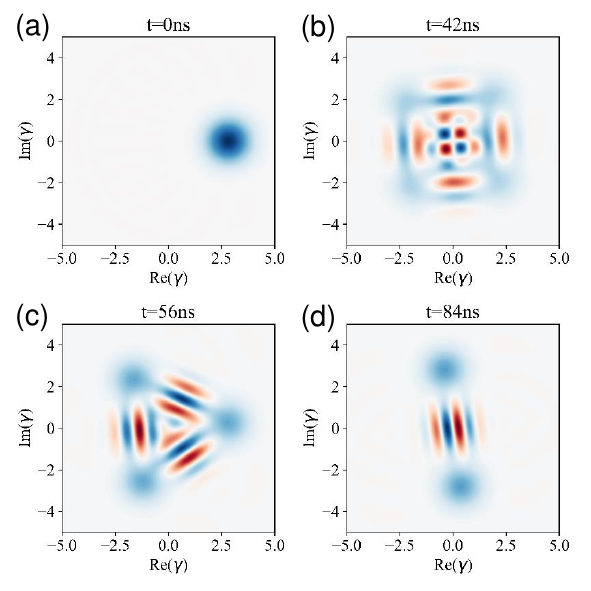}
\caption{\label{fig:8} Numerical Wigner functions of the Kerr Cat state when $\alpha=2$. (a) When $t=0$, it is a coherent state. (b) When $t=42ns$, the coherent state evolves into a four-body Cat state. (c) When $t=56ns$, the coherent state evolves into a three-body Cat state. (d) When $t=84ns$, the coherent state evolves into a two-body Cat state.}
\end{figure}

\begin{figure}
\includegraphics[width=0.48\textwidth]{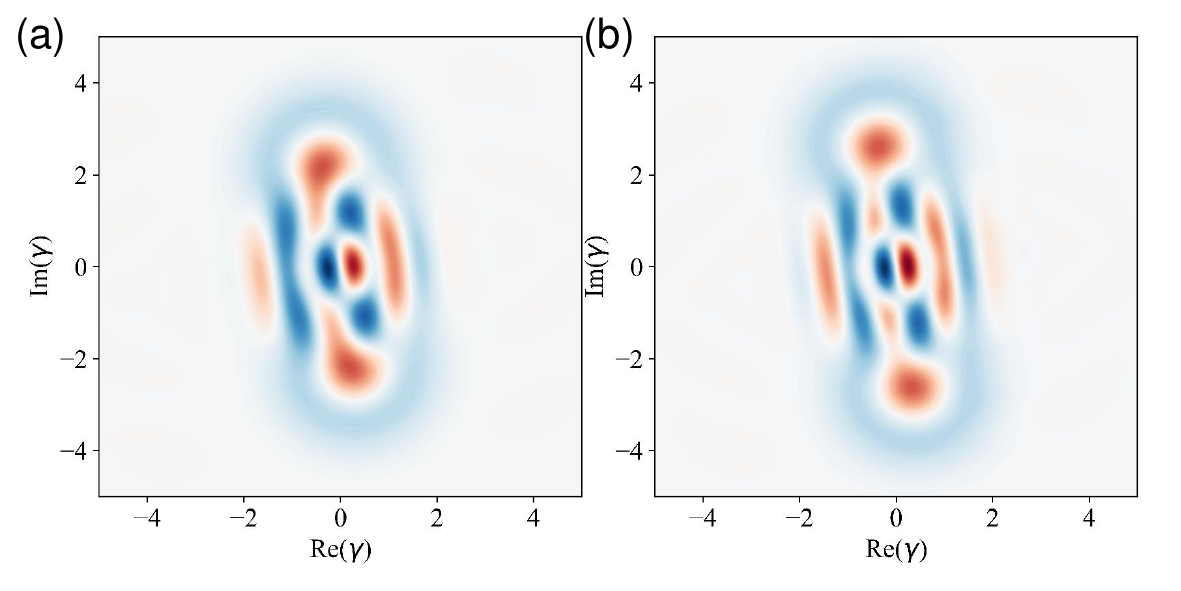}
\caption{\label{fig:9} Quantum correlation of two-body Cat states with different $\alpha$ (a)Quantum correlation of two-body Cat states for photon numbers $n=2$ and $n=3$. (b)Quantum correlation of two-body Cat states for photon numbers $n=3$ and $n=4$.}
\end{figure}

In our simulation, the flux bias pulse in the coupler is turned off at this point so that transmon and fluxonium can interact with each other. Wigner tomography can accomplish by applying a qubit to drive the transmon and a readout pulse to drive the readout resonator. The pulse sequence of the Kerr Cat state as shown in Appendix C. We chose $\alpha$=2 to simulate the evolution of the coherent state to the two-body Cat state. We use a non-linear Kerr Hamiltonian to simulate the generation of Cat states. The flux bias pulse in SQUID sets the Kerr non-linearity coefficient to $K=-5.95MHz$, the corresponding evolution time is $t=168ns$. When the widths of the flux bias pulses are $t_1=84ns$, $t_2=56ns$, and $t_3=42ns$, respectively, we achieve the two-, three-, and four-body Cat states, and the numerical simulation results of the Wigner tomography are shown in Fig.~\ref{fig:8}.

In addition, we also calculate the quantum correlation of the two-body Cat states with different photon numbers $n$, where $\alpha=\sqrt n$. The quantum correlation of multi-body Cat state can be written as

\begin{equation}
\begin{split}
C_m(\gamma)=W_m^{\alpha=\sqrt{n+1}}(\gamma)-W_m^{\alpha=\sqrt{n}}(\gamma),
\end{split}
\end{equation}

Where $W_m^{\alpha=\sqrt{n}}(\gamma)$ is the Wigner function when multi-body Cat state at $\alpha=\sqrt{n}$, and $n$ is the number of photons. The quantum systems with different photon numbers $n$ can be obtained by varying the displacement $\alpha$ of the pump pulse. As shown in Fig.~\ref{fig:9}, figure 9 (a) is the two-body Cat state correlation when $n=2$ and $n=3$, and figure 9 (b) is the two-body Cat state correlation when $n=3$ and $n=4$. The quantum correlation of the two-body Cat state is similar to the shape of the Cat state, but it shows the characteristics of layer upon layer of nesting.

\section{CONCLUSION}

In conclusion, we propose a tunable coupling scheme based on fluxonium-transmon-transmon (FTT), which is used to realize a high fidelity two-qubit gates with fixed frequency fluxonium and transmon. In this system, both qubits and coupler are capacitively coupled, and the coupler is a frequency tunable transmon qubit. By decoupling the coupler from the system, the effective coupling strength can be easily adjusted to zero to close the net coupling between qubits. We study the performance of this scheme by simulating the general single-qubit $(X_{\frac{\pi}{2}})$ gate
and two-qubit $(iSWAP)$ gate. We simulate single-qubit gates and two-qubit gates (iSWAP) on circuits with FTT structures, respectively. Single-qubit gates can achieve 99.99$\%$ fidelity within $15ns$. Without decoherence, two-qubit gates can achieve 99.95$\%$ fidelity within $103ns$. 

FTT structure is a feasible scheme in quantum circuit structure. By selecting appropriate circuit devices to form the qubit frequency and the coupling strength, which can operate the quantum gate. The capacitive coupling of different types of superconducting qubits can provide a new approach for achieving scalable and large-scale superconducting quantum processors. This asymmetric structure not only generates high-fidelity two-qubit gates, but also generates Schrödinger Cat states in fluxonium. By adjusting the nonlinear Kerr coefficient of fluxonium, the multi-body Schrödinger Cat state can be obtained in Wigner tomography. Compared with ordinary superconducting circuits, we consider the nonlinear Kerr effect of fluxonium and realize the Cat states, and analyze the quantum correlation of two-body Cat states. Our simulation shows that the energy level gap between fluxonium at $5.7GHz$ and transmon at $6.4GHz$ is very loose, but the coupling still stable, which is very effective for multi-level structures. Our method can effectively solve the problem of frequency-crowding and form a high fidelity quantum gate, at the same time produce a Schrödinger Cat state in the qubits.

\section{Acknowledgement}
Financial support from the project funded by the State Key Laboratory of Quantum Optics and Quantum Optics Devices, Shanxi University, Shanxi, China (Grants No.KF202004 and No.KF202205).

\appendix

\section{Circuit Hamiltonian And Quantization}

The circuit diagram for realizing our adjustable coupling scheme as shown in Fig.~\ref{fig:1}(a), which contains two qubits: qubit 1 (blue) and qubit 2 (green) and a coupler (red). Qubit1 is a fixed frequency fluxonium, qubit2 is a fixed frequency transmon, and the coupler is a frequency adjustable transmon. Low frequency fluxonium consists of a capacitor, an inductor and a Josephson junction.

We choose the flux node $\phi_i$ corresponds to node $i$ in the graph as the generalized coordinates of the system. We can write out the Lagrange function of the circuit by using the flux node $\phi_i$:
\begin{equation}
\begin{split}
T=&\frac{1}{2}[C_1\dot\phi_1^2+C_c\dot\phi_c^2+C_2\dot\phi_2^2+C_{1c}(\dot\phi_1-\dot\phi_c)^2\\&+C_{2c}(\dot\phi_2-\dot\phi_c)^2+C_{12}(\dot\phi_1-\dot\phi_2)^2],
\end{split}
\end{equation}

\begin{equation}
\begin{split}
U=&-\beta E_{J_1}\cos\left(\frac{2\pi}{\Phi_0}\phi_1\right)-E_{J_1}\cos\left(\frac{2\pi}{\Phi_0}(\phi_{ext}-\phi_1)\right)\\&-E_{J_2}\cos\left(\frac{2\pi}{\Phi_0}\phi_2\right)-E_{J_c}\cos\left(\frac{2\pi}{\Phi_0}\phi_c\right)+\frac{\phi_1^2}{2L_1},
\end{split}
\end{equation}

\begin{equation}
\begin{split}
L=T-U,
\end{split}
\end{equation}

Where $\Phi_0=h/2e$ is the flux quantum. Writing out the kinetic energy part in matrix form:
\begin{equation}
\begin{split}
T=\frac{1}{2}\dot{\vec{\phi}}^TC\dot{\vec{\phi}},
\end{split}
\end{equation}

The capacitance matrix is:
\begin{equation}
\begin{split}
C=\left[\begin{array}{ccc}
C_1+C_{1c}+C_{12} & -C_{1c} & -C_{12}\\
-C_{1c} & C_c+C_{1c}+C_{2c} & -C_{2c}\\
-C_{12} & -C_{2c} & C_2+C_{2c}+C_{12}
\end{array}\right]
\end{split}
\end{equation}

The inverse capacitance matrix is:
\begin{equation}
\begin{split}
C^{-1}=\left[\begin{array}{ccc}
\frac{1}{C_1} & \frac{C_{1c}}{C_1C_c} & \frac{C_{12}+(C_{1c}C_{2c})/C_c}{C_1C_2}\\
\frac{C_{1c}}{C_1C_c} & \frac{1}{C_c} & \frac{C_{2c}}{C_2C_c}\\
\frac{C_{12}+(C_{1c}C_{2c})/C_c}{C_1C_2} & \frac{C_{2c}}{C_2C_c} & \frac{1}{C_2}
\end{array}\right]
\end{split}
\end{equation}

The Hamiltonian of the system can be obtained:
\begin{equation}
\begin{split}
\hat{H}=&\sum_{i}{q_i\dot\phi_i}-L=\frac{1}{2}\vec{q}^T[C^{-1}]\vec{q}+U\\&=\frac{q_1^2}{2C_1}-\beta E_{J_1}\cos\left(\frac{2\pi}{\Phi_0}\phi_1\right)-E_{J_1}\cos\left(\frac{2\pi}{\Phi_0}(\phi_{ext}-\phi_1)\right)\\&+\frac{\phi_1^2}{2L_1}+\frac{q_2^2}{2C_2}-E_{J_2}\cos\left(\frac{2\pi}{\Phi_0}\phi_2\right)+\frac{q_c^2}{2C_c}-E_{J_c}\cos\left(\frac{2\pi}{\Phi_0}\phi_c\right)\\&+\frac{C_{1c}}{C_1C_c}q_1q_c+\frac{C_{2c}}{C_2C_c}q_2q_c+\left(\frac{C_{12}}{C_1C_2}+\frac{C_{1c}C_{2c}}{C_1C_2C_c}\right)q_{1}q_2,
\end{split}
\end{equation}

By using the canonical quantization, the quantized Hamiltonian can be rewritten as
\begin{equation}
\begin{split}
\hat{H}=&4E_{c_1}\hat{n}_1^2-\beta E_{J_1}\cos\varphi_1-E_{J_1}\cos\left(\varphi_{ext}-\varphi_1\right)+\frac{1}{2}E_{L_1}\varphi_1^2\\&+4E_{c_2}\hat{n}_2^2-E_{J_2}\cos\varphi_2+4E_{cc}\hat{n}_c^2-E_{J_c}\cos\varphi_c\\&+8\frac{C_{1c}}{\sqrt{C_1C_c}}\sqrt{E_{c1}E_{cc}}(\hat{n}_1\hat{n}_c)+8\frac{C_{2c}}{\sqrt{C_2C_c}}\sqrt{E_{c2}E_{cc}}(\hat{n}_2\hat{n}_c)\\&+8(1+\eta)\frac{C_{12}}{\sqrt{C_1C_2}}\sqrt{E_{c1}E_{c2}}(\hat{n}_1\hat{n}_2),
\end{split}
\end{equation}

Where $\eta=C_{1c}C_{2c}/C_{12}C_c$ is the parameter factor, $\hat{n}_\lambda=\hat{q}_\lambda/2e$ is the Cooper pair number operator, $E_L=\left(\frac{\Phi_0}{2\pi}\right)^2/L$ is the inductance energy of the corresponding mode and $E_{C_m}=e^2/2C_m$ is the charging energy of the corresponding mode.

In the fluxonium and transmon coupling regime, we substitute the creation (annihilation) operator into the Hamiltonian

\begin{equation}
\begin{split}
\varphi=\left(\frac{2E_C}{E_J}\right)^\frac{1}{4}(\hat{a}^\dagger+\hat{a}),
\end{split}
\end{equation}

\begin{equation}
\begin{split}
\hat{n}=\frac{i}{2}\left(\frac{E_J}{2E_C}\right)^\frac{1}{4}(\hat{a}^\dagger-\hat{a}),
\end{split}
\end{equation}

The Hamiltonian of the system becomes ($\hbar=1$):

\begin{equation}
\begin{split}
\hat{H}=\hat{H}_{q_1}+\hat{H}_{q_2}+\hat{H}_c+\hat{H}_{1_c}+\hat{H}_{2_c}+\hat{H}_{12},
\end{split}
\end{equation}

\begin{equation}
\begin{split}
\hat{H}_{q_1}=\sqrt{8E_{J_1}E_{C_1}c_2}\hat{a}_1^\dagger\hat{a}_1+g_3(\hat{a}_1+\hat{a}_1^\dagger)^3+g_4(\hat{a}_1+\hat{a}_1^\dagger)^4,
\end{split}
\end{equation}

\begin{equation}
\begin{split}
\hat{H}_m=\sqrt{8E_{J_m}E_{C_m}}\hat{a}_m^\dagger\hat{a}_m-\frac{E_{C_m}}{2}\hat{a}_m^\dagger\hat{a}_m^\dagger\hat{a}_m\hat{a}_m, m=q_2,c,
\end{split}
\end{equation}

\begin{equation}
\begin{split}
\hat{H}_{j_c}=g_j(\hat{a}_j^\dagger\hat{a}_c+\hat{a}_j\hat{a}_c^\dagger-\hat{a}_j^\dagger\hat{a}_c^\dagger-\hat{a}_j\hat{a}_c), j=1,2,
\end{split}
\end{equation}

\begin{equation}
\begin{split}
\hat{H}_{12}=g_{12}(\hat{a}_1^\dagger\hat{a}_2+\hat{a}_1\hat{a}_2^\dagger-\hat{a}_1^\dagger\hat{a}_2^\dagger-\hat{a}_1\hat{a}_2),
\end{split}
\end{equation}

Where $\hat{a}_m(\hat{a}_m^\dagger)$ corresponds to the creation (annihilation) operator of the corresponding mode, Taylor coefficient $c_i=\frac{1}{E_J}\frac{d^iU_{fluxonium}}{d\varphi_1^i}|_{\varphi_{min}}$, $g_3(g_4)$ is the coupling strength of the three(four)-wave mixing,

\begin{equation}
\begin{split}
\hbar g_3=\frac{1}{6}\frac{c_3}{c_2}\sqrt{E_{C_2}\hbar\omega_{q_1}},
\end{split}
\end{equation}

\begin{equation}
\begin{split}
\hbar g_4=\frac{1}{12}\frac{c_4}{c_2}E_{C_2},
\end{split}
\end{equation}

We approximate looks fluxonium as a harmonic oscillator, the nonlinear kerr coefficient $K$ can be calculated as the dispersion of the transition frequencies between the neighboring energy levels, 

\begin{equation}
\begin{split}
\hbar K=\frac{d^2E_n}{dn^2}=12\hbar(g_4-\frac{5g_3^2}{\omega_{q_1}}),
\end{split}
\end{equation}

The coupling strength is

\begin{equation}
\begin{split}
g_j=\frac{1}{2}\frac{C_{j_c}}{\sqrt{C_jC_c}}\sqrt{\omega_j\omega_c},j=1,2,
\end{split}
\end{equation}

\begin{equation}
\begin{split}
g_{12}=\frac{1}{2}(1+\eta)\frac{C_{12}}{\sqrt{C_1C_2}}\sqrt{\omega_1\omega_2}.
\end{split}
\end{equation}

$g_j$ is the coupling strength of qubit-coupler, and $g_{12}$ is the coupling strength of qubit-qubit. The equation not only retains the interaction term, but also retains the conter-rotating term. The contribution of double excitation interaction is also significant in the dispersion regime where the coupler frequency is much greater than that of the qubit frequency.

\section{Schrieffer-Wolff Transformation}

In order to decouple, we apply the Schrieffer-Wolff transformation
\begin{equation}
\begin{split}
\hat{U}=Exp[\sum_{j=1,2}\frac{g_j}{\Delta_j}(\hat{a}_j^\dagger\hat{a}_c-\hat{a}_j\hat{a}_c^\dagger)],
\end{split}
\end{equation}

Where $\Delta_j=\omega_j-\omega_c$, we expand $\tilde{U}\hat{H}\tilde{U}^\dagger$ in the second order. We finally obtain the effectively Hamiltonian:

\begin{equation}
\begin{split}
\hat{\tilde{H}}=&\tilde{\omega}_1\hat{a}_1^\dagger\hat{a}_1+K\hat{a}_1^\dagger\hat{a}_1^\dagger\hat{a}_1\hat{a}_1+\tilde{\omega}_2\hat{a}_2^\dagger\hat{a}_2\\&+\frac{\tilde{\alpha_2}}{2}\hat{a}_2^\dagger\hat{a}_2^\dagger\hat{a}_2\hat{a}_2+\tilde{g}(\hat{a}_1^\dagger\hat{a}_2+\hat{a}_1\hat{a}_2^\dagger),
\end{split}
\end{equation}

where 

\begin{equation}
\begin{split}
\tilde{\omega}_1\approx\omega_1+\frac{g_1^2}{\Delta_1},
\end{split}
\end{equation}

\begin{equation}
\begin{split}
\tilde{\alpha}_1\approx-E_{J_1},
\end{split}
\end{equation}

\begin{equation}
\begin{split}
\tilde{\omega}_2\approx\omega_2+\frac{g_2^2}{\Delta_2},
\end{split}
\end{equation}

\begin{equation}
\begin{split}
\tilde{\alpha}_2\approx-E_{C_2},
\end{split}
\end{equation}

\begin{equation}
\begin{split}
\tilde{g}\approx\frac{g_1g_2}{2}\left(\frac{1}{\Delta_1}+\frac{1}{\Delta_2}\right)+g_{12}.
\end{split}
\label{equ:B7}
\end{equation}

In Eq.\ref{equ:B7}, we consider the effect of rotation term and obtain the effective coupling strength between fluxonium and transmon. The computational states $|100\rangle$ and $|001\rangle$ can be exchanged through the Jaynes-Cummings interaction\cite{shore1993} $(\hat{a}_j^\dagger\hat{a}_c+\hat{a}_j\hat{a}_c^\dagger)$.

\section{Pulse Sequence}

\begin{figure}
\includegraphics[width=0.48\textwidth]{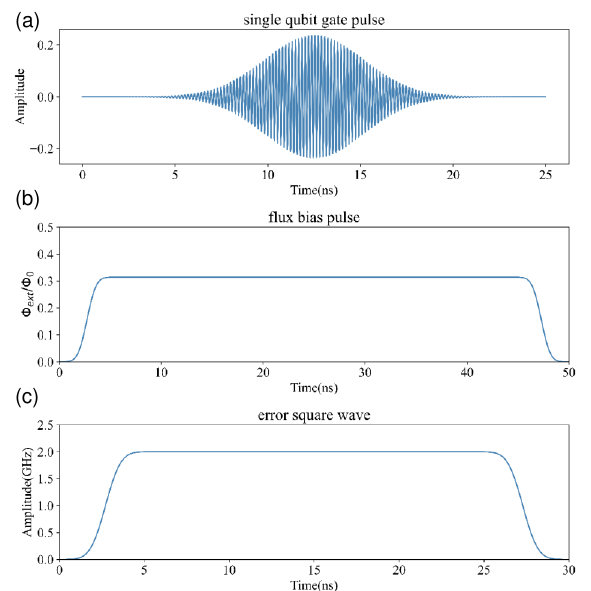}
\caption{\label{fig:10} (a) Single-qubit gate pulses act on qubits. (b) Flux bias pulse act on coupler. (c) Error type square waves act on qubits.}
\end{figure}

\begin{figure*}
\centering
\includegraphics[width=1.0\textwidth]{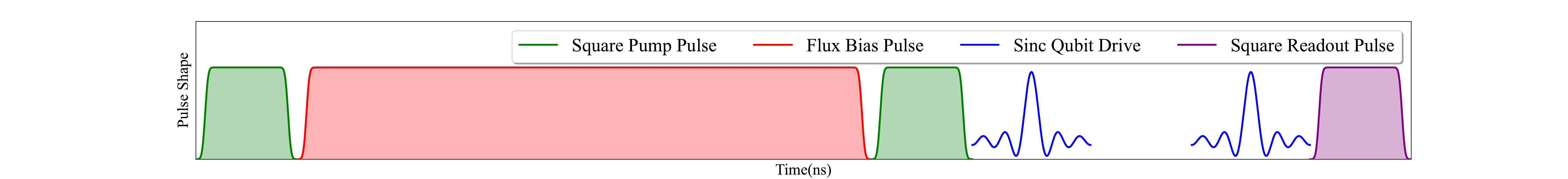}
\caption{\label{fig:11} A pulse sequence generated by a entanglement state. The square pump pulses(the green pulses), the flux bias pulses(the red pulses) acts on fluxonium, the qubit driving pulses (the blue pulses) acts on transmon, and the readout pulses (the purple pulses) acts on the readout resonator.}
\end{figure*}

The pulse sequence used for quantum gate operation as shown in Fig.~\ref{fig:10}, and the pulse sequence used in entanglement state generation and numerical simulation as shown in Fig.~\ref{fig:11}. 

A single-qubit gate pulse consists of a microwave source provide a stable drive pulse. An arbitrary wave generator (AWG) provide a base-band pulse. The frequency of the driving pulse is $\omega_{L_O}$, and the envelope of the base-band pulse is $s(t)$, and the frequency is $\omega_{AWG}$. The IQ-mixer combines the two signals to produce a new signal with a waveform of $s(t)$ and a frequency $\omega_d=\omega_{L_O}\pm\omega_{AWG}$.

The rectangular flux pulse passing through the flux bias line to the coupler's Josephson junction loop with a magnitude of a multiple of the superconducting flux quantum $\Phi_0$. During the flux bias, the effective coupling strength can be adjusted to a suitable frequency.

Error type square waves acts on the qubits through an XY line, the amplitude of which is the energy level difference between the two qubits. The effect is to compensate for the energy level difference between the qubits, so that the two qubits can reach a resonance state. During the pulse, the two-qubit gate works normally.

In Fig.~\ref{fig:11}, the first green pulse is a square pump pulse acts on fluxonium, which causes a displacement $D(\alpha)$ make the vacuum state becomes coherent state $|\alpha\rangle$.

The first red pulse is the flux bias pulse acts on SQUID in fluxonium, which can adjust the nonlinear Kerr coefficient to an appropriate strength, and the duration of the pulse can be determined by the nonlinear Kerr coefficient, and the Kerr Cat state generated in fluxonium under the evolution of nonlinear Hamiltonian.

The next four pulses together form the pulse sequence of the Wigner tomography. The second green pulse brings a displacement $D(-\gamma)$. Here, we apply two $\pi/2$ pulses (blue pulse) into transmon. Using the time interval $\pi/(g_2^2/\Delta_2)$ between the two pulses, this sequence can be considered as a parity measurement, where the state $|g\rangle(|e\rangle)$ of transmon corresponds to the measurement probability $P=-1(1)$. Considering the frequency shift of transmon, we employ a function of $sinc(t)=\sin(t)/t$ to cover a larger spectrum range uniformly. The state of transmon can be measured by the final readout pulse (the purple pulse) applied to a readout resonator capacitively coupled to transmon.

\nocite{*}
\bibliography{FFT}

\end{document}